\documentclass[smallextended]{svjour3}       

\smartqed  
\usepackage{graphicx}
 \journalname{Experimental Astronomy}

\newcommand{\SII}{[S~{\sc ii}]}
\newcommand{\OIII}{[O~{\sc iii}]}
\newcommand{\NII}{[N~{\sc ii}]}
\newcommand{\HII}{H~{\sc ii}}
\newcommand{\Ha}{H$\alpha$}

\newcommand{\ergs}{\,\mbox{erg}\,\mbox{s}^{-1}\,\mbox{cm}^{-2}\,\mbox{arcsec}^{-2}}

\newcommand{\SIIHa}{[S~{\sc ii}]/H$\alpha$}
\newcommand{\NIIHa}{[N~{\sc ii}]/H$\alpha$}
\newcommand{\OIIIHb}{[O~{\sc iii}]/H$\beta$}
\def\mangal{MaNGaL\,}
\def\aj{AJ}
\def\apj{ApJ}

\def\aap{A\&A}
\def\mnras{MNRAS}

\def\apjl{ApJLett}
\def\pasa{PASA}
\def\pasp{PASP}

\begin{document}

\title{Mapper of Narrow Galaxy Lines  (MaNGaL): new tunable filter  imager for Caucasian   telescopes
}

\titlerunning{Mapper of Narrow Galaxy Lines  (MaNGaL)}        

\author{Alexei Moiseev$^{1,2}$      \and Alexander~Perepelitsyn$^1$ \and Dmitry Oparin$^1$
}

\authorrunning{Alexei Moiseev et al.} 

\institute{ A. Moiseev \at
              $^1$Special Astrophysical Observatory, Russian Academy of Sciences, Nizhny Arkhyz 369167, Russia \\
              \email{moisav@gmail.com}           
              ORCID: 0000-0002-0507-9307 
              \and
              $^2$Lomonosov Moscow State University, Sternberg Astronomical Institute, Universitetsky pr. 13, Moscow 119234, Russia\\
}

\date{Received: date / Accepted: date}

\maketitle

\begin{abstract}
We described the design and operation principles of a new tunable-filter  photometer  developed  for the 1-m  telescope of the  Special Astrophysical Observatory of the Russian Academy of Sciences  and the 2.5-m telescope of the Sternberg Astronomical Institute of the Moscow State  University. The instrument is mounted on the scanning Fabry-Perot interferometer operating in the tunable-filter mode in the spectral range of 460-800 nm with a typical spectral resolution of about 1.3 nm. It allows one to create images of galactic and extragalactic nebulae in the emission lines  having different excitation conditions and to carry out diagnostics of the gas ionization state. The main steps of observations, data calibration, and reduction are  illustrated  by examples of  different emission-line objects: galactic  \HII{} regions, planetary nebulae, active galaxies with extended filaments, starburst galaxies, and Perseus galaxy cluster.

\end{abstract}

\section{Introduction}
\label{intro}

\begin{figure*}
	\includegraphics[width=\textwidth]{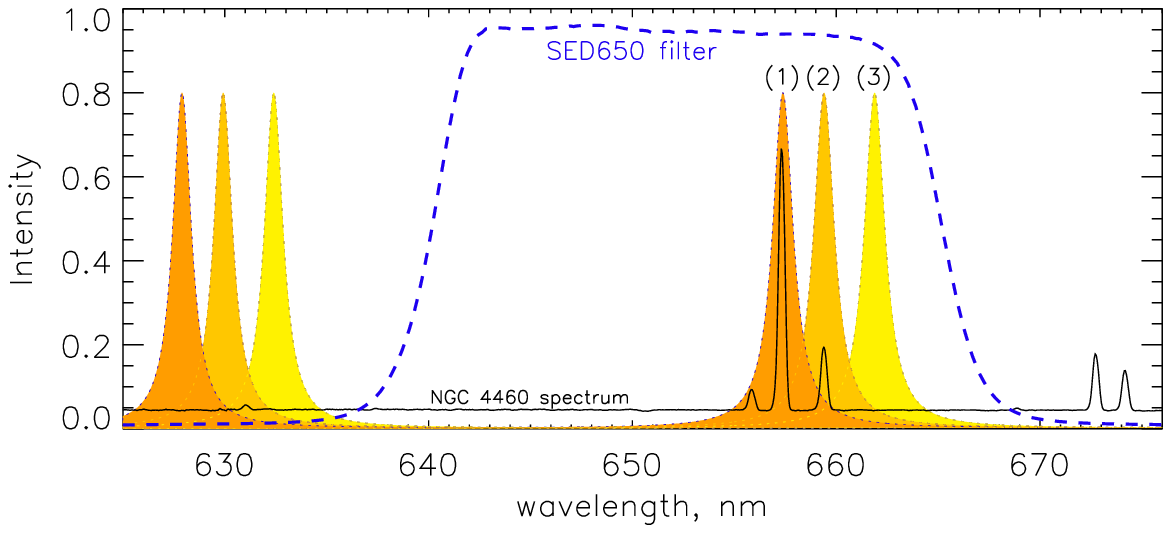}
	\caption{Tunable filter operating principle. The regions  filled with various tones of orange show the transmission profiles of the FPI tuned for observations in the \Ha\ (1) and \NII$\lambda6583$ (2) emission lines  and in the continuum (3). The blue dashed line shows the transmission curve of the medium-band filter which isolates only one transmission peak of the interferometer. The parameters of FPI and medium-band filter are similar to those used in the MaNGaL; the integrated spectrum of the NGC~4460 starburst galaxy was taken from \cite{Moiseev2010}.
	}
	\label{fig:ngc4460}       
\end{figure*}

Tunable filter (TF) imaging systems based on low-order scanning Fabry-Perot interferometers (FPIs) have a long history of astronomical applications related to the study of extended emission-line targets: galactic and extragalactic nebulae, solar system objects \cite{Jockers1992,Bland1998,Jones2002}.  The main idea of observations  are well described in the above references and also illustrated in Fig.~\ref{fig:ngc4460}.   If the  gap between FPI plates is small and  corresponds to the interference orders $n=10-30$, then it is easy to attain the  FWHM of the instrumental profile $\delta\lambda=1-2$~nm. Since the  distance between neighbouring interference orders (interfringe) $\Delta\lambda=\lambda/n$, we can cut the desired  transmission peaks  by the medium-band filter with a typical bandwidth of about 15--30 nm.
The peak transmission central wavelength (CWL) can be  switched between the desired emission line and neighboring continuum using  a  piezoelectrically-tuned and servo-stabilized FPI; the redshift/systemic velocity of the studied objects can be taken into account. 

 This technique  allows us to make  accurate continuum subtraction  to produce pure line images and precisely distinguish the neighbouring emission lines (the \Ha{} and \NII$\lambda6548,6583$  lines for the ionization state and metallicity diagnostics or \SII$\lambda6717,6731$ lines for measurements of the electron density $n_e$).  The last is usually impossible in the  `traditional' medium- or narrow-band filter  imaging.  From the other hand,  the TF observations give a significantly large field of view (FOV) compared to the  integral-field spectroscopic observations:  several and even tens of arcminutes  with  large telescopes, whereas the integral-field system MUSE/VLT, the most powerful today, has only $1'$ FOV\cite{MUSE}. Several TF systems are in operation  at  large telescopes like the  6.5-m Magellan \cite{MMTF}, 8.2-m  SUBARU \cite{Sugai2010}, or 10.4-m GTC \cite{OSIRIS}.

TFs have great research potential in observations with medium-sized telescopes and the Taurus Tunable Filter \cite{Bland1998} at the 3.9-m Anglo-Australian Telescope was the first to demonstrate this. The obtained results included the discovery of compact galaxies emitting in the \Ha line at the $z\sim0.4$ redshift \cite{Jones2001}, mapping of the extended gas around quasars\cite{Shopbell1999}, the study of meteorological processes in the atmospheres of brown dwarfs using time variability at the TiO absorption feature \cite{Tinney1999} and many other interesting findings.
Nevertheless, the TF technique is not very popular now, because it  is much more complicated than traditional filter  imaging; moreover, some differences from `an ideal monochromator' are also known. We list the problems according to \cite{Jones2002}:
 \begin{enumerate}
 	\item The instrumental transmission Airy profile is more triangular rather than rectangular or Gaussian, and this complicates the flux calibration and separation of neighbouring lines.  
 	\item Using the  medium-band blocking filter  limits the operating spectral range. 
 	\item The size of the quasi-monochromatic region (Jacquinot spot) is limited, because the FPI phase changes across FOV.
 \end{enumerate}

In this paper, we describe the TF photometer developed for  observations at  Russian medium-sized  telescopes, where problem (3) is partially  solved by placing  the FPI in the convergent beam instead of the collimated  one. Also, using  a large set of  modern high-transparency blocking filters allows  us to reduce  problem (2) of the spectral range selection.

\begin{figure*}
	\includegraphics[width=\textwidth]{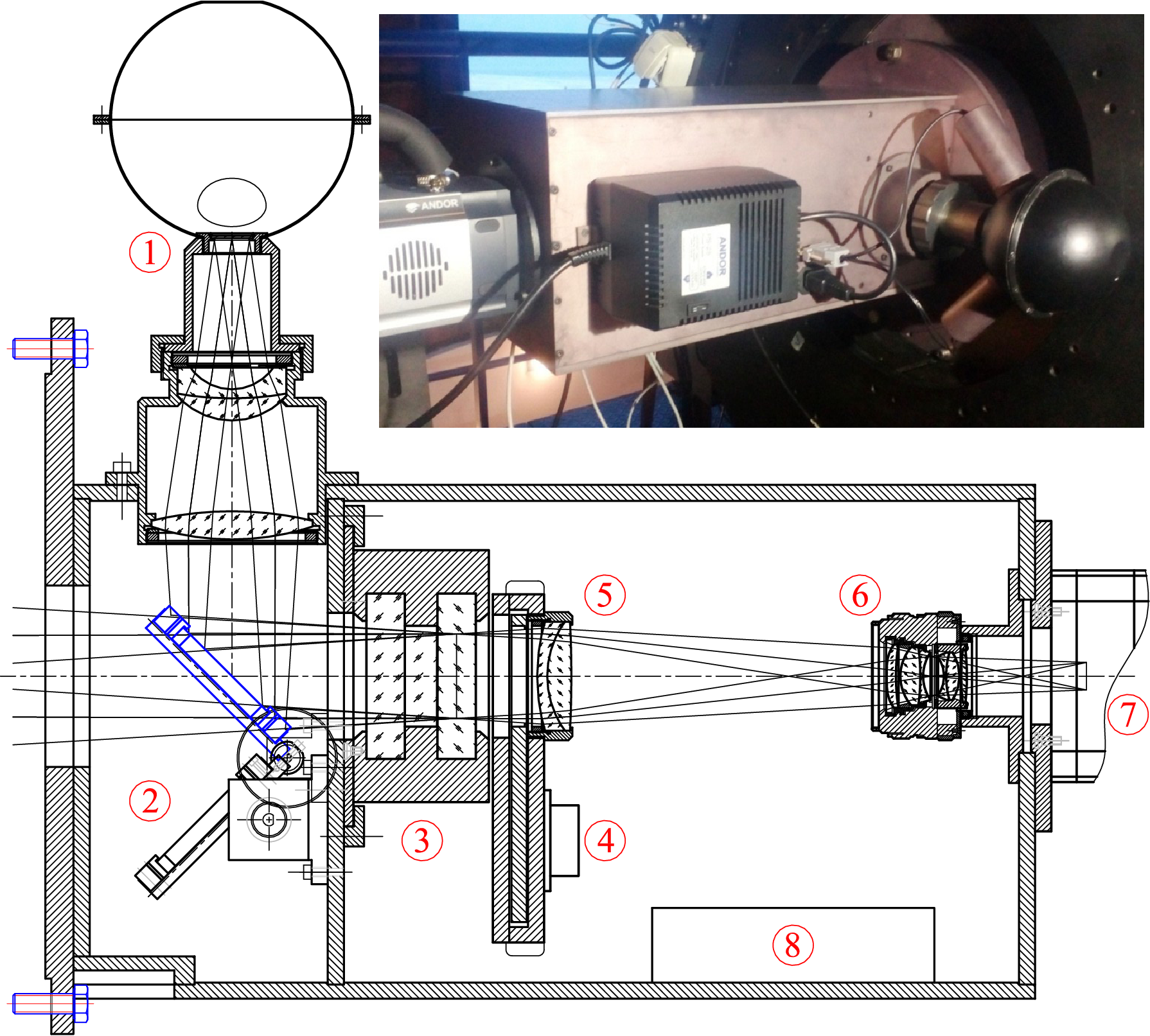}
	\caption{ \mangal optomechanical layout and photo of the device in the Nasmyth-2 focus at the 2.5-m SAI MSU telescope. (1) -- the calibration unit; (2) -- the diagonal mirror; (3) -- the scanning FPI; (4) -- the medium-band filters wheel; (5) -- the field lens; (6) -- the photo lens; (7) -- the CCD camera; (8) -- the control computer.}
	\label{fig:optics}      
\end{figure*}

\section{Design of the instrument}
\label{sec:1}

\subsection{Optical and mechanical layout}

\begin{figure*}
	\includegraphics[width=0.5\textwidth]{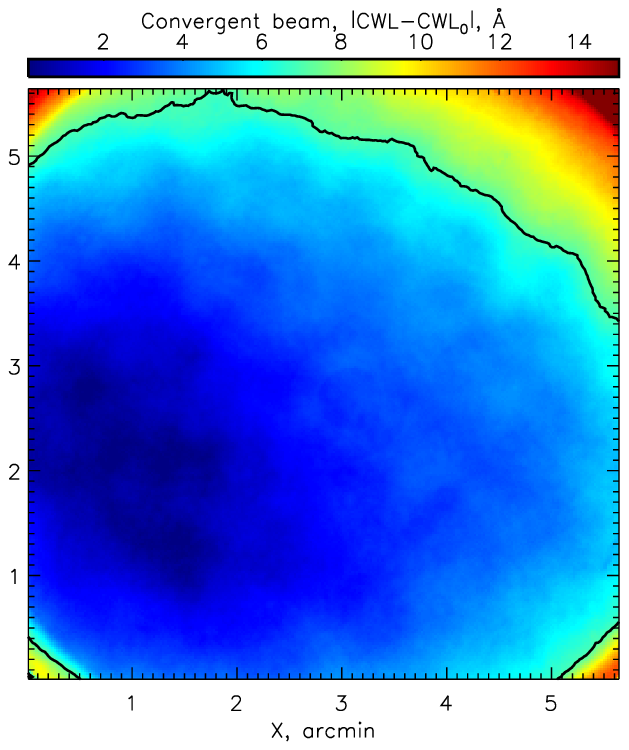}
	\includegraphics[width=0.5\textwidth]{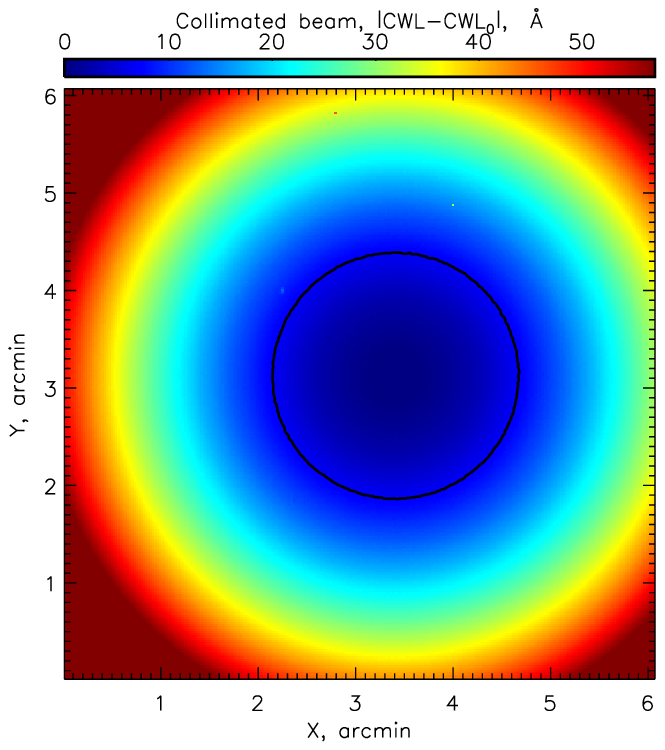}
	\caption{CWL variations across the similar field-of-view for the same low-order FPI  in  different optical systems. The black thick line marks the Jacquinot spot borders ($\Delta$CWL$<\frac{1}{2}\delta\lambda$). 
		Direct measurements using the calibration lamp spectrum: FPI in the convergent beam in  MaNGaL (left) and in the collimated beam  in SCORPIO-2 focal reducer \cite{SCORPIO2}  (right).
	}
	\label{fig:spot}       
\end{figure*}

The Mapper of Narrow Galaxy Lines  (MaNGaL\footnote{`Mangal' is a Caucasian and Middle-East barbeque.}) was  developed and   manufactured in the Special Astrophysical Observatory of the  Russian Academy of Sciences (SAO RAS).  The instrument optical scheme (Fig.~\ref{fig:optics}) consists of a two-component  achromatic field lens and a six-component anastigmatic photographic lens `Helios-44M-7' (F/2). In contrast to  the `classical' optical layout having a TF in the collimated beam \cite{Jones2002,MMTF}, \mangal is an afocal reducer with the FPI in the convergent  beam as was proposed by Courtes \cite{Courtes1960}. This arrangement provides a significantly larger size of a central monochromatic region that is crucial in studying the extended targets. Figure~\ref{fig:spot} shows the variations of the instrumental CWL (`phase map', see \cite{Moiseev2002ifp,Jones2002}) for the same low-order FPI in two different optical schemes with the similar camera focal ratio: in the collimated beam in the SCORPIO-2 focal reducer and in    \mangal. The figure evidences that  the region, where the peak  CWL changes less than $0.5\delta\lambda$,  covers $\sim90\%$ of the \mangal FOV, that is  significantly larger than in the collimated beam. Moreover, here we use a more rigid criteria than the standard definition of the Jacquinot spot: the peak wavelength variation over the field  does not exceed  $\sqrt{2}\delta\lambda$ \cite{Jacquinot1954,Jones2002}. With this standard definition, the monochromatic spot covers the whole \mangal FOV. It is important to note that this optical scheme type works only with low-order interferometers, because  the instrumental finesse ($\Delta\lambda/\delta\lambda$) degrades very fast with the interference order in   the same  convergent beam.

The mechanical part of the device also includes  a USB motorized filter wheel  with a changeable 5-position holder for 50-mm diameter filters produced by Edmund Optics\footnote{https://www.edmundoptics.com/f/motorized-filter-wheels/13430/}, and a rotating diagonal mirror  which directs the light from the calibration unit. The telecentric optics of the calibration unit produces an image  of the area illuminated by calibration lamps in the Ulbricht integrating sphere to the entrance of the focal reducer. The focal ratio of the calibration beam is about $F/8$.    The integration sphere can be  illuminated by the He-Ne-Ar filled lamp  (the Soviet vintage vacuum stabilitron SG3S) to calibrate the wavelength scale and the filament lamp to produce the spectral  flat field. 

To select the desired spectral range around the FPI transmission peak, we use  hard coated bandpass filters also produced by Edmund Optics. The main  set of the filters with the 25-nm bandwidth is similar to that described in  \cite{Dodonov2017}. These filters uniformly cover the 460--750-nm wavelength interval, their transmission curve has an almost rectangular shape with a maximum throughput of $\sim95\%$ (see the example in Fig.~\ref{fig:ngc4460}). Also, we use several filters with the 10-nm bandwidth which cover the spectral regions between the 25-nm filter profiles.

\begin{figure*}
	\includegraphics[width=0.5\textwidth]{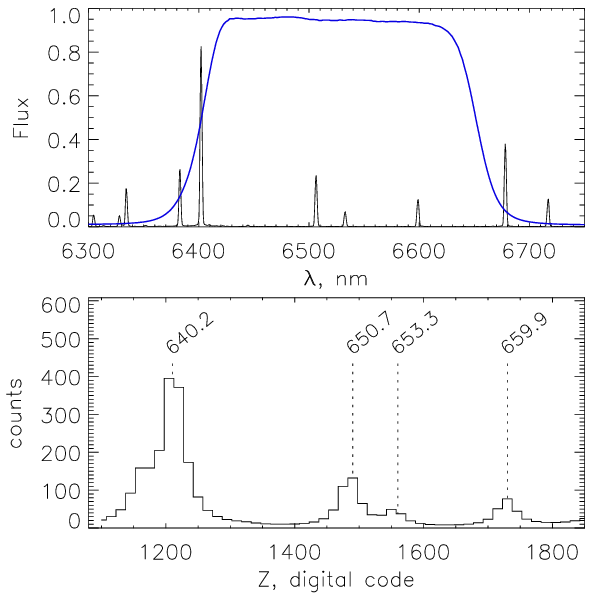}
	\includegraphics[width=0.5\textwidth]{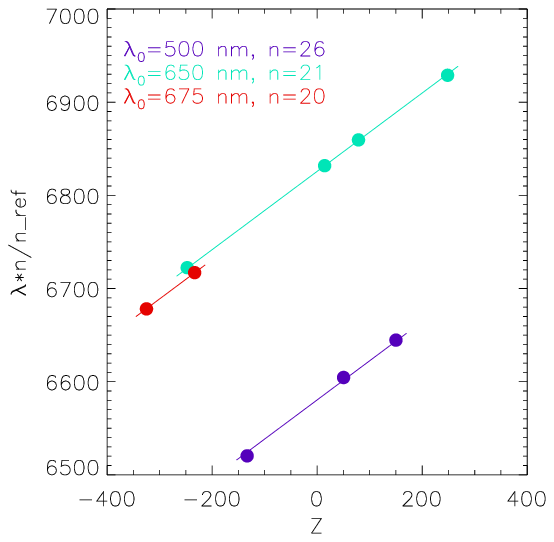}
	\caption{Left:  the spectrum of the He-Ne-Ar calibration lamp, the blue line shows the transmission curve of the bandpass filter selected the 25 nm region around  the \Ha{} line  (top) and  the selected spectrum scanned with the FPI (bottom). The brightest calibration lines are marked. Right: the wavelength calibration relation, different colors correspond to different spectral ranges, the CWLs of the bandpass filters are labeled, $n_{ref}=21$ .}
	\label{fig:calib}      
\end{figure*}

The AVR ATtiny2313 microprocessor controls the diagonal mirror and calibration lamps, whereas the compact MR3253S-00F industrial computer  is used for  full control of all optical and mechanical units including the CCD camera and scanning FPI controller CS-100 (see below). The control interface for working  with the local microprocessor, scanning FPI, the filter wheel, and the CCD acquisition system is written in the IDL data language.

\subsection{CCD}
As a  detector, \mangal uses the iKon-M 934  camera   manufactured by Andor\footnote{http://www.andor.com/}  based on a  $1K\times1K$ back-illuminated CCD optimized for the visual spectral range (QE$\ge90\%$  in the 500--700 nm wavelength range). The camera is connected to the control computer via the USB-2 interface. The   Peltier cooling system of the CCD together with the industrial chiller provide a   working temperature of $-100^\circ$C    (the coolant is the $50\%$ water-alcohol mixture). The CCD read noise is 2.5 or 6  $\mathrm{e^-}$, it depends on the read-out speed. The pixel scale and  FOV provided by this detector at  different telescopes are listed in Tab.~\ref{tab:1}.

%
\begin{table}
	\caption{Main parameters of MaNGaL at the SAO RAS and SAI MSU telescopes}
	\label{tab:1}       
	\begin{tabular}{lll}
		\hline\noalign{\smallskip}
		& 1-m SAO RAS & 2.5-m SAI MSU  \\
		&   (Cassegrain, F/13) &  (Nasmyth-2, F/8)\\
		\noalign{\smallskip}\hline\noalign{\smallskip}
		Total  focal ratio & F/5.26 & F/3.25 \\
		Field of view  & $8.7'$& $5.6'$ \\
		Pixel scale             & 0.51$''$ & 0.33$''$ \\
		Spectral range            & \multicolumn{2}{c}{460--750~nm} \\
		Spectral resolution       & \multicolumn{2}{c}{1.0--1.6~nm} \\
			\noalign{\smallskip}\hline
	\end{tabular}
\end{table}

\subsection{Scanning FPI}

The ET-50  piezoelectric FPI (made by IC Optical Systems Ltd\footnote{http://www.icopticalsystems.com/}) with the appropriate characteristics was bought by SAO RAS within the project for modernization of the equipment of the 6-m SAO RAS telescope (`the BTA Unique Scientific Instrument'). To effectively use the FPI, when it was not involved in observations at the 6-m SAO RAS telescope, it was decided to use it  as a guest TF for   medium-sized telescopes. The working order of interference is $n\approx20$ (at $\lambda=656.3$ nm) that corresponds  to an interferometer nominal cavity spacing of about $10\,\mu$m. The FPI reflector coating is optimized for work in the spectral range of 460--750 nm with a finesse of about $\Delta\lambda/\delta\lambda=16-27$. The corresponding typical  spectral resolution is $\delta\lambda\approx1.3\pm0.3$ nm. Using the digital controller  CS-100, we can change  the gap between the interferometer plates and scanning  spectra with an accuracy in  wavelength of $~0.04$~nm.

\begin{figure*}
	\centerline{
		\includegraphics[width=0.5\textwidth]{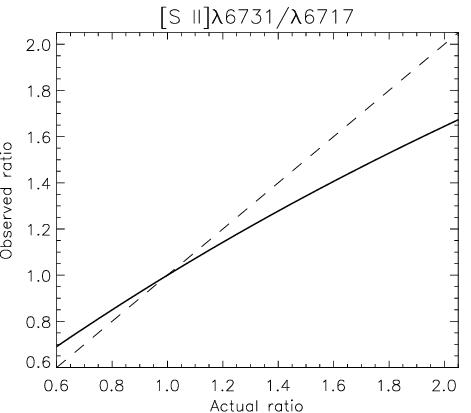}
		\includegraphics[width=0.5\textwidth]{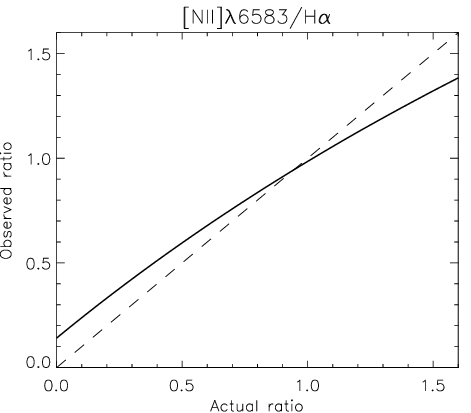}
	}
	\caption{Calibration relations between the actual  ($r$) and the observed  ($r_{obs}$) line flux ratios for the neighbouring  emission line pairs: \SII$\lambda6731/\lambda6717$ (left)  and \NII/\Ha{} (right). The Lorentz instrumental profile with $\delta\lambda=1.2$~nm was accepted. The solid line shows the calibration relation, whereas the dashed line corresponds to $r=r_{obs}$.  
	}
	\label{fig:ratios}       
\end{figure*}

\section{Wavelength calibration}
\label{sec:wave}

In the scanning piezoelectrically-tuned interferometer, the plate spacing is changed   by setting  the digital value $Z_s$ to the CS-100 controller. In this case, the  relationship between this value, wavelength, and interference order can be written as \cite{Moiseev2002ifp,Jones2002}: 

\begin{equation}
n\lambda =A+BZ_s.
\label{eq1}
\end{equation}

The current order in each wavelength with the same gap   is related to some `referred order' $n_{ref}$ as: 

\begin{equation}
n\lambda =n_{ref}\lambda_{ref},
\label{eq2}
\end{equation}
where we accepted $\lambda_{ref}=656.3$~nm (the \Ha{} line). Figure~\ref{fig:calib} shows the example of the calibration lamp emission lines selected by the blocking filter and scanned with our FPI. We have 2--4 unblended calibration lines   in each 25-nm blocking filter, therefore,  we can determine three constants ($n_{ref}$, $A$, $B$) from (\ref{eq1}) and (\ref{eq2}). However, in the case of a relatively small gap (10~$\mu$m in our case), the wavelength-dependent phase change in the reflections between optical coatings on the inner plate surfaces could be important \cite{Jones2002,MMTF}. In practice, it means that $A$ is not a constant,  but depends on wavelength  $A=A(\lambda)$. Therefore,  we use a linear fitting of  the relations  ($n\lambda$ vs. $Z_s$) with  slope $B$, but with different $A(\lambda)$ in  different blocking filters. Figure~\ref{fig:calib}  (right) shows the example of this fitting for the center of  FOV. The shift of the filter CWL for each point across  FOV is also determined from the scanning of the spectra of the same calibration lamp (see Fig.~\ref{fig:spot}). The position of lines is determined using the fitting by the Lorentz contour which gives a good approximation  of the FPI  instrumental profile \cite{Jones2002,Moiseev2002ifp}:
\begin{equation}
L(\lambda-\lambda_0) = \frac{1}{1+\left( \frac{2(\lambda-\lambda_0)}{\delta\lambda} \right)^2},
\label{eq_lor}
\end{equation}
where  $\lambda_0$ is a line position.


\section{Data reduction}

\subsection{Preliminary reduction}  
The  \mangal data reduction are similar to that  used for  `standard' CCD direct-image  frames. It includes: calculation and subtraction of the mean bias (with this CCD, the dark current is insufficient 
for   typical exposures of 10-20 min), flat-field correction, and alignment of monochromatic images using the images of stars.  Then we combine  individual  exposures at the same wavelength with  the cosmic ray hits removal based on the standard sigma-clipping technique.

The flat-field frames are illuminated by the continuum spectra lamp  using the same passband filter and the same scanning  $Z_S$ values, as those of the object frames. Thanks to the relatively large  Jacquinot spot, we  avoid   the complicated process of the removal of rings produced by the airglow night-sky   emission (see \cite{Moiseev2002ifp,Jones2002}). Instead, in the most  cases, we can accept the uniform distribution of the night-sky emission   across the field.

The continuum emission is removed from the images in the lines  with multiplication of the continuum frames to the coefficient  $C_{cont}$ to minimize the flux residuals in the field stars. In practice, if we  alternate the continuum and line exposures (see Sec.~\ref{sec:methods}) and normalize  the frames to  the exposure time, then  this coefficient is  affected only by the blocking filter transmission variations and  $C_{cont}\approx1.00\pm0.1$, because we try to avoid observations near the  wings of filter transmission curve.

\subsection{Flux calibration}

Calibration of CCD counts to the flux in  physical units [$\mbox{erg}\,\mbox{s}^{-1}\,\mbox{cm}^{-2}$] is performed    
according to the algorithm and equations given in \cite{Jones2002}.  Briefly, we observe standard stars with the same settings (the blocking filter and  $Z_S$) as the object  frames. The expected flux from a standard star is calculated as a convolution of the tabulated star spectrum and FPI instrumental profile, that is the Lorentz profile with $\delta\lambda$ obtained from the fitting of calibration lamp lines, see (\ref{eq_lor}). We observe the spectrophotometric standard stars from the ESO list\footnote{https://www.eso.org/sci/observing/tools/standards/spectra.html} which is mainly based on the data from \cite{Oke1990}. Correction  for atmospheric extinction is performed with the extinction curve  for the SAO RAS site   \cite{Kartasheva1978}.

The detected spectrum is a product of a convolution of an actual spectrum with the instrumental profile $L(\lambda)$. If two emission lines separated in  wavelength as $\Delta\lambda_{01}$ has  fluxes $F_1$, $F_2$ and their ratio $r_{01}=F_1/F_0$ then our tunable filter measurement gives a smoothed observed ratio:

\begin{equation}
r_{obs}=\frac{F_1+F_0L(\Delta\lambda_{01})}{F_0+F_1L(\Delta\lambda_{01})}
=\frac{r_{01}+L(\Delta\lambda_{01})}{1+r_{01}L(\Delta\lambda_{01})}.
\end{equation}

Therefore:

\begin{equation}
r_{01}=\frac{r_{obs}-L(\Delta\lambda_{01})}{1-
	r_{obs}L(\Delta\lambda_{01})}.
\label{eq_SII}
\end{equation}

The left-hand panel in Fig.~\ref{fig:ratios} demonstrates  calibration relation (\ref{eq_SII}) for the case of the  \SII{} doublet line  ratio used for $n_e$ estimations. 

In the case of three lines, the last equation can be written as:  

\begin{equation}
r_{01}=\frac{r_{obs}-L(\Delta\lambda_{01})}{1+r_{12}*L(\Delta\lambda_{12})-
	r_{obs}[L(\Delta\lambda_{01})+C*L(\Delta\lambda_{02})]},
\label{eq_Ha}
\end{equation}
where  $\Delta\lambda_{ij}=(\lambda_i-\lambda_j)$. The   Figure~\ref{fig:ratios} (right) shows this relation for the well-known \NII$\lambda6583$/\Ha{} line ratio, where  index `2' corresponds to the \NII$\lambda6548$ line, and $r_{12}=1/3$.   

Calibration relations  (\ref{eq_SII}) and  (\ref{eq_Ha}) should be taken into account, when the corresponding line ratio maps were are created.

\begin{figure*}
	\includegraphics[width=\textwidth]{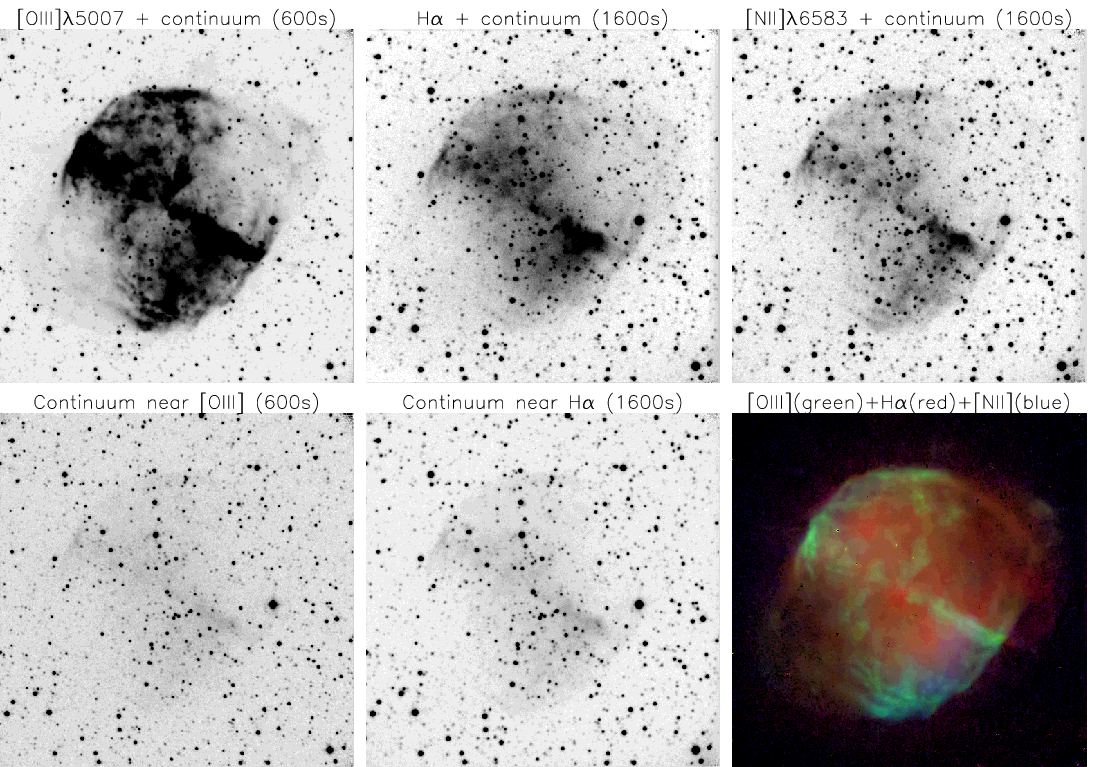}
	\caption{Images of the planetary nebula  NGC 6853 (Messier~27)  in the case of the FPI transmission peak centered at the emission lines and the continuum, as well as the color composite image of the nebula in the \OIII$\lambda5007$, \Ha, and \NII$\lambda6583$ lines after subtracting the stellar continuum. The total exposures are written in brackets.}
	\label{fig:M27}       
\end{figure*}

\begin{figure*}
	\includegraphics[width=\textwidth]{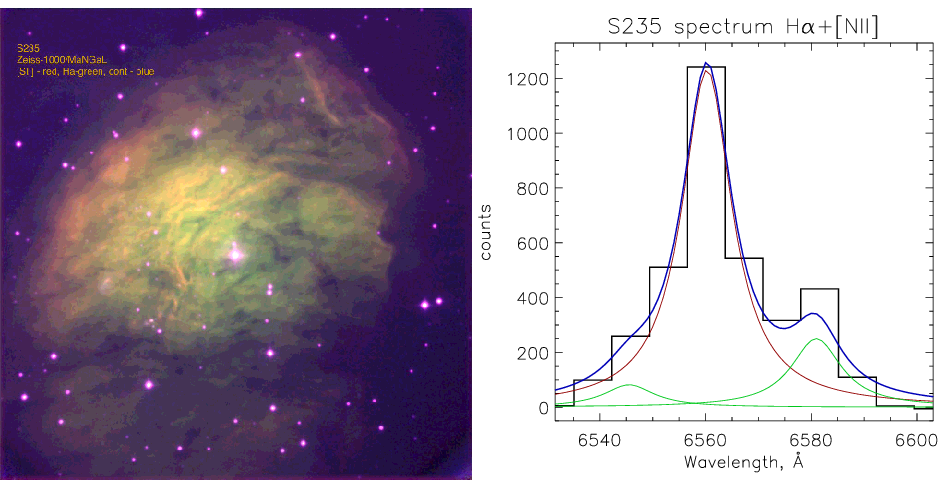}
	\caption{Observations of the Sh2-235 galactic HII region at the 1-m telescope. Left: the color composite image in the \SII\ (red), \Ha{} (green), and continuum (blue). Right: the example of the Lorentzian fitting of the  \Ha+\NII\ spectrum of the nebulae obtained with \mangal in the scanning mode. }
	\label{fig:S235}       
\end{figure*}

\begin{figure*}
	\includegraphics[width=0.5\textwidth]{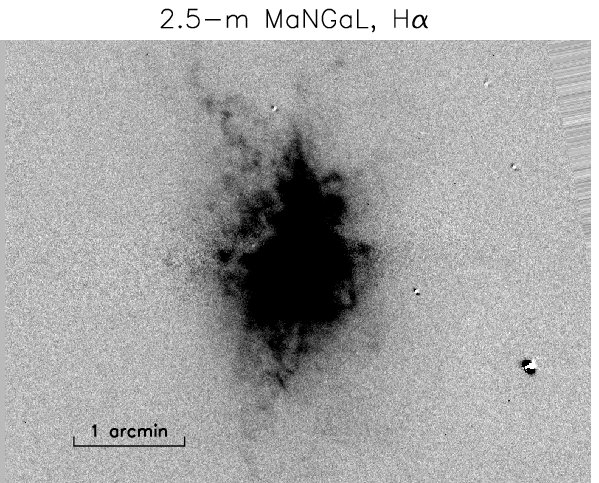}
	\includegraphics[width=0.5\textwidth]{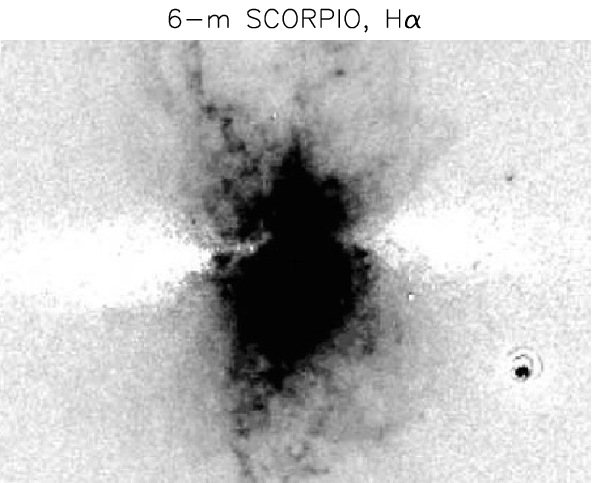}
	\caption{Continuum-subtracted \Ha\ images  of the galactic wind in NGC 3034 (Messier~82)  observed at the 2.5-m SAI MSU telescope  with the tunable filter in \mangal (left) and the 6-m telescope data from \cite{Karachentsev2007} obtained with classical medium-band filters technique (right).}
	\label{fig:M82 }      
\end{figure*}

\section{Observations}

\label{sec:methods}

\subsection{Main principles}
Before  and after an observation night, we perform  the \mangal calibration: scanning the wavelength range  in the selected bandpass filters for both He-Ne-Ar and flat-field lamps. Also, we  monitor the FPI stability using a brief scan of the spectral   region around the brightest calibration line in the desired filter just before  the object exposure. It allows us to know the current  $A$ value in equation (\ref{eq1}). 

The main mode of \mangal observations is the \textit{`TF mode'}, when  the CWL is tuned first to the emission line (taking into account the systemic velocity of the target and heliocentric correction), and then to the neighboring continuum (shifted by 3--5~nm).  In the case of observations of the line doublets (\Ha+\NII, \SII, etc.) the sequence can  be: `line1--line2--continuum'. Also it is possible to observe the red and blue continuum, for example:    `red continuum--line1--line2--blue continuum'. This cycle is  repeated, which averages the variations of seeing and  atmospheric extinction.

In the \textit{`scanning FPI mode'}, we quickly scan the wavelength regions around the emission line with typical   CWL increments of 0.5-0.8~nm ($\sim0.5\delta\lambda$). In this case, we are  able to obtain  the low-resolution data cube to study the ionized gas kinematics and collect emission spanning a large velocity range.

\subsection{SAO RAS 1-m  telescope}

We developed \mangal  as a guest instrument for two medium-sized telescopes located in the Northen Caucasus region of the Russian Federation: the 1-m telescope of SAO RAS and 2.5-m telescope of the Sternberg Astronomical Institute, Lomonosov Moscow State  University (SAI MSU). The parameters of the instrument at  different telescopes are listed in  Tab.~\ref{tab:1}.

Figure~\ref{fig:M27} demonstrates the results of the first light \mangal observations of the planetary nebulae Messier~27 at the Cassegrain focus of the 1-m telescope Zeiss-1000' in September 2017: the images in  the emission lines and the continuum, as well as the color composite image of the nebula in the \OIII, \Ha, and \NII{} lines after subtracting the stellar continuum.  The relatively large field of view in the Cassegrain focus of Zeiss-1000 is convenient for observations of galaxies having a large angular diameter and Galactic \HII{} regions. A good   example is the Sh2-235 region of star formation, its emission-line  composite  image  restored from \mangal observations  is shown in Fig.~\ref{fig:S235}. Also, this figure shows the  spectrum in the \Ha+\NII{} emission lines of this nebulae obtained in the scanning mode of \mangal. The detailed description and analysis of  these observations are given in \cite{Kirsanova2019}.

\subsection{SAI MSU 2.5-m telescope}

More compact but fainter targets were observed  at the 2.5-m alt-azimuthal telescope of the SAI MSU    on Mt. Shatdzhatmaz \cite{Kornilov2014}.  \mangal was mounted at the rotating stage  of the Nasmyth-2 focus. An advantage of a tunable filter mapping comparing with a conventional technique of medium-band imaging is illustrated by Fig.~\ref{fig:M82 }. Here, the left-hand panel presents the \Ha{} emission line image of a prototype galactic wing nebulae in the local galaxy Messier~82 observed with \mangal at the 2.5-m telescope, whereas the right-hand   panel shows the \Ha+\NII{} view on the same region taken with at the 6-m telescope with  the SCORPIO focal reducer (filter FWHMs were 7.5~nm for the emission line and 17--20 nm for the blue and red continuum). Both \mangal and SCORPIO data  sets have a similar angular scale (0.33$''$ and 0.36$''$ per px) and  total exposure (500 and 600 sec correspondingly), therefore the detected signal is significantly higher in the 6-m telescope observations, because  telescope aperture and total transparency of the  telescope+device system, are larger. However, using of \mangal allowed us to subtract very accurately the  stellar host galaxy disc in contrast with the oversubtracted horizontal region in the SCORPIO data. It allows us to study weak gas  emission in the region,  where the  underlying  stellar continuum is dominated in the total spectrum.

\begin{figure*}
	 \includegraphics[width=0.5\textwidth]{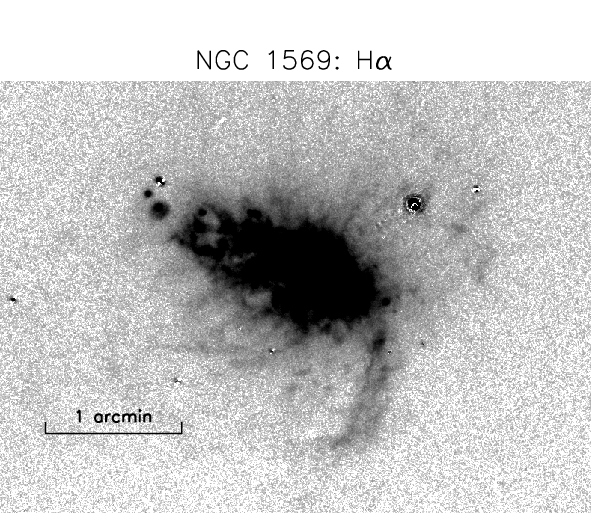}
	 \includegraphics[width=0.5\textwidth]{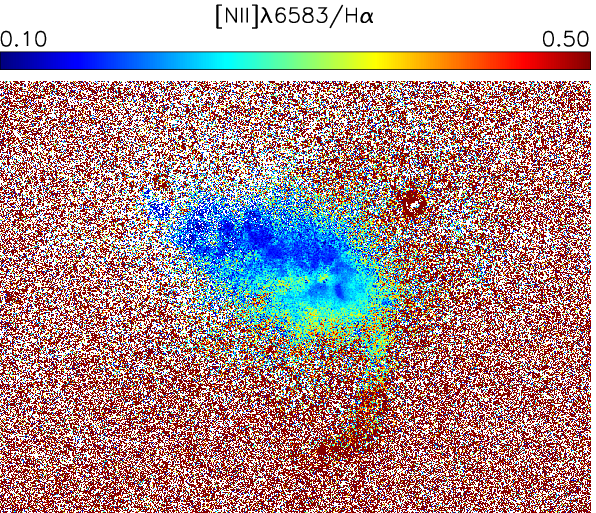}
	   \includegraphics[width=\textwidth]{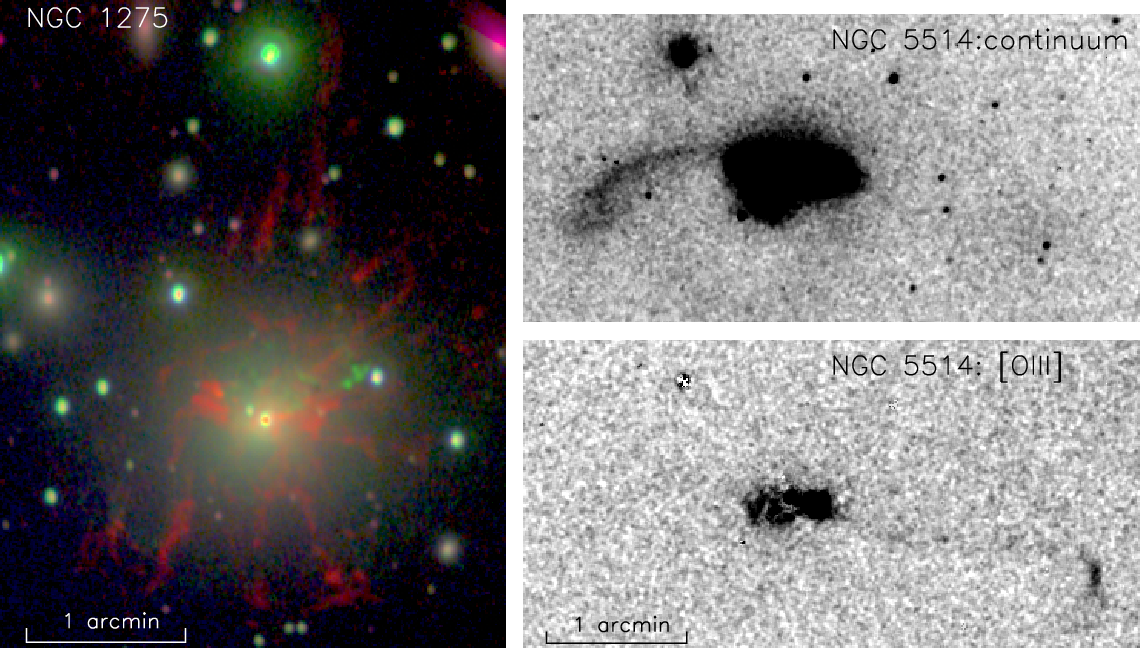}
		 \includegraphics[width=0.33\textwidth]{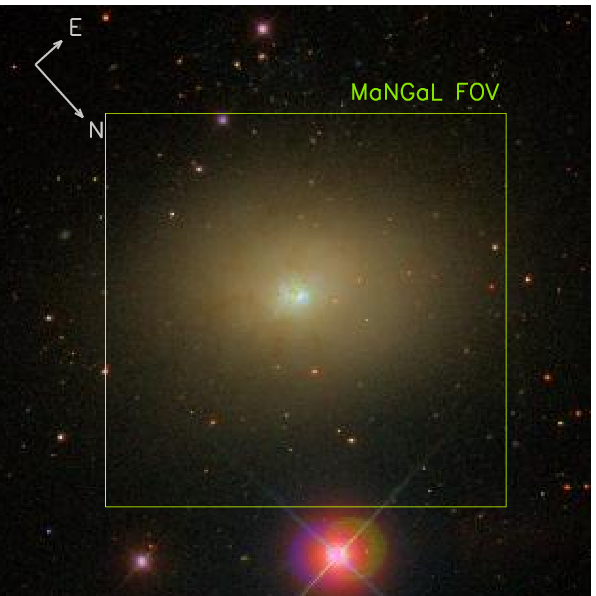}
\includegraphics[width=0.66\textwidth]{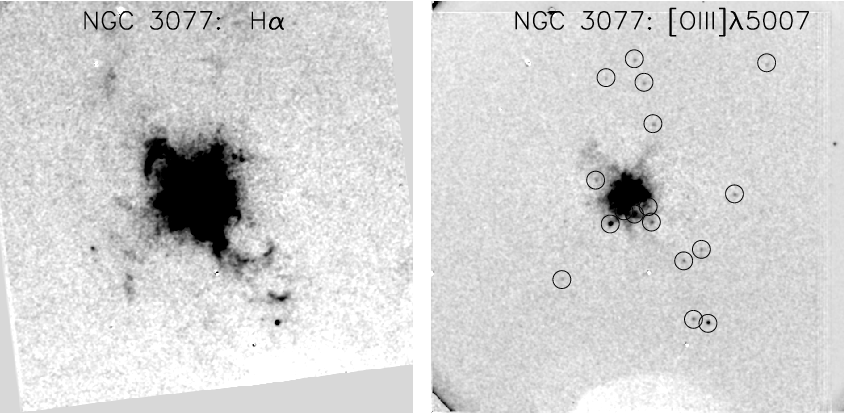}
	\caption{Various emission-line  objects observed with \mangal  at the 2.5-m SAI MSU telescope, from top to bottom: the \Ha-image of the galactic wind in NGC 1569 (left) and the map of the \NII$\lambda$6583/\Ha{} ratio (right); the central galaxy of the Persieus cluster NGC~1275 (left) seen in the \NII{} (red channel), and \OIII{} (green) emission lines  and the images of NGC~5514 interaction pair in the continuum and in the  \OIII$\lambda5007$ emission (right); SDSS image of the nearby dwarf galaxy NGC~3077, the \mangal $5.6'$ field of view is marked (left), also the emission line images in \Ha{} (center) and \OIII$\lambda5007$ (right) are shown. Circles mark the compact emission knots -- PNe candidates.
}
	\label{fig:objects}      
\end{figure*}

Figure~\ref{fig:objects}  demonstrates the capabilities of \mangal{} at  the 2.5-m telescope for study of the ionized gas in extragalactic objects of different types:

\begin{itemize}
    \item NGC~1569. Another example of a local starburst galaxy, where the numerous  supernova explosions formed a system of suberbubbles and galactic wind outflow. The \NIIHa{} ratio map shows that the young stars dominate in the gas ionization in the central region (\NIIHa$\approx0.16$), whereas the  shock ionization due to a galactic wind is important in the external emission filaments (those \NIIHa$\ge0.5$ are in a good agreement with previous spectral observations (see Fig.~6 in \cite{Heckman1995}).
    \item NGC~1275. The \NII{} image reveals the well-known system of the ionized gas filaments extended around this central galaxy of the Perseus  cluster. The \OIII{} emission in the filament is very weak, it mainly comes from a   high-velocity system of gaseous clouds identified as a disrupted foreground galaxy northwest of the NGC~1275 nucleus  \cite{Meaburn1989,Boroson1990}.
   \item NGC~5514.  A major-merger galaxy pair with a large tidal tail in the  continuum image. The \OIII{} map reveals for the first time the extended gaseous clouds  ionized by the active nucleus at a distance of more than 60 kpc. This galaxy has been selected for \mangal observations as external ionized clouds candidate in the TELPERION survey \cite{Knese2020}. 
\item NGC~3077.  The dust lanes in optical images of this dwarf starburst galaxy in the  Messier~81 group are related to the  ionized gas filaments \cite{Walter2002}. Our \Ha{} image (total exposure of 900 sec) reveals  all structural features appeared in one of deepest images in the literature \cite{Karachentsev2007}. In contrast to the \Ha{} image, the \OIII{} emission map revealed about twenty compact emission objects. The observed \OIII/\Ha{} ratio allows us to  consider the most of them as planetary nebulae (PNe)   candidates.     
\end{itemize}

Typical  detection sensitivity  of  faint diffuse emission has been estimated in nights with photometric atmospheric conditions. At  the 1-m SAO RAS telescope, the \Ha{} emission was detected with the signal-to-noise ratio $S/N=3$ at the $1\times10^{-16}\ergs$ surface-brightness level (an exposure of 1200 sec, the $1\times1$ binning mode, 26 Jan, 2018, the Sh2-235 nebulae). The corresponding value in 2.5-m telescope observations was  $2\times10^{-16}\ergs$ (an exposure  of 900 sec, the $2\times2$ binning mode, 10 Apr, 2018, the NGC~3077 galaxy).

\begin{figure*}
	\includegraphics[width=\textwidth]{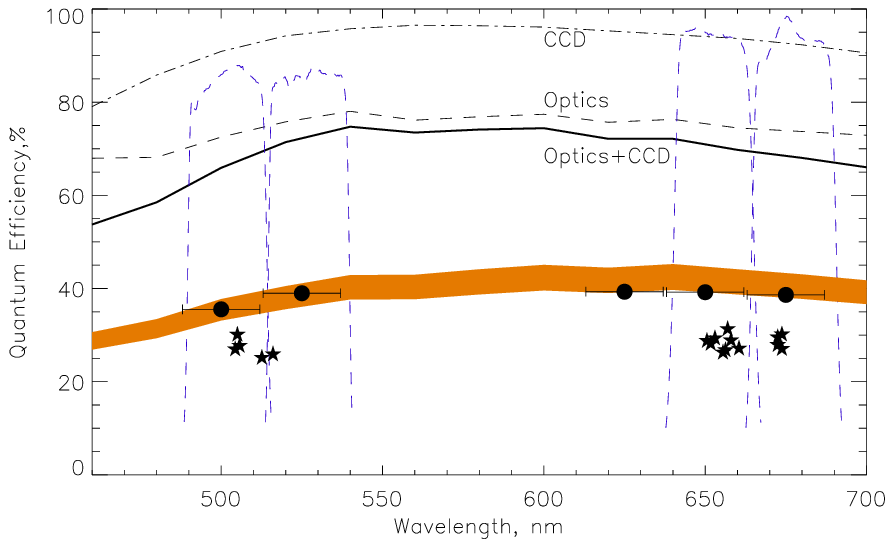}
	\caption{QE of \mangal at the 2.5-m telescope. Different black lines mark  the transparency of different \mangal components according to the labels. The middle-band filters transmission curves are shown by  blue lines. The orange-filled region is the total optics+telescope efficiency, its width corresponds to the mirrors' reflecion uncertainty. The black symbols  show the QE measurements  using spectrophotometric standards observed without FPI (circles) and with the  tunable filter (stars).} 
	\label{fig:qe}       
\end{figure*}

Estimations of the \mangal$+2.5$-m telescope quantum efficiency (QE) and contribution of different components to the total  throughput are presented in Fig.~\ref{fig:qe}. The transmission curves of the optics and filters were measured in the SAO RAS laboratory; the CCD QE is given according to the  manufacturer's data. We use the direct estimation of  three telescope mirrors (M1, M2 and M3) provided by the observatory staff. The prediction of the total QE telescope$+$\mangal in the direct-image mode (without FPI)  has demonstrated a very good agreement with observations of the standard star observed in a night with  photometric atmospheric conditions in Nov 2017. The total QE in the TF mode is   smaller, because it includes the light lost in FPI (observations were carried out in Apr 2018). The value of total QE$=25-30\%$ is similar to that obtained with the Taurus Tunable Filter at the 3.9-m Anglo-Australian telescope \cite{Jones2002}. According to our measurements, the  \mangal efficiency at the 1-m Ziess-1000 telescope has the same value, because its two mirrors in the Cassiegrain  focus give the  same losses of light  as three  2.5-m telescope mirrors in its Nasmyth focus.

\section{Conclusion}

In this paper, we described the design  of the \mangal{} tunable-filter photometer and the first experience in the study of emission-line  extended targets with the SAO RAS and SAI MSU medium-sized telescopes. 
Sciencific results published or submitted  so far include the discovery  of cross-ionization of  gaseous discs   by the active nuclei hosted in  each companion galaxies in the interacting pair UGC 6081   \cite{Keel2019}, the discovery of a gaseous spiral structure in the lenticular galaxy NGC~4143 \cite{Silchenko2020}, the study of the gas ionization state in the emission filaments of NGC~3077 \cite{Oparin2020}, the reconstruction of a spatial structure of  the \HII{} region Sh2-235 \cite{Kirsanova2019}.  The most effective usage of \mangal consists in    observation in a couple of neighbouring emission lines with different mechanism of excitation  having  a common point for the continuum  (\Ha{}+\NII{}, e.g.). However, it also can be used to create diagnostic diagrams based on several line ratios, like \OIIIHb{} vs. \SIIHa{} \cite{Oparin2020}. Also, \mangal{} is a  very effective instrument in combination  with other optical spectroscopic data obtained with long-slit spectrographs or high-resolution  scanning Fabry-Perot interferometers, as it was shown in our works listed above.

\begin{acknowledgements}
This study was supported by the Russian Science Foundation, project no. 17-12-01335 `Ionized gas in galaxy discs and beyond the optical radius'. Observations conducted with the  telescopes of the Special Astrophysical Observatory of the Russian Academy of Sciences carried out with the financial support of the Ministry of Science and Higher Education of the Russian Federation (including agreement No. 05.619.21.0016, project ID RFMEFI61919X0016). The work of 2.5-m telescope is  supported by  the Program of development of M.V. Lomonosov Moscow State University).
The Andor detector was purchased within the framework of the Russian Science Foundation grant no.14-22-00041. The authors are thankful to  Victor Afanasiev and Vladimir Amirkhanyan for their help and  fruitful comments on the \mangal design,  Marya Burlak, Victor Komarov, Andrey Tatarnikov, Nikolay Shatskii, Victor Senik, and Olga Voziakova for their help and assistance in 1-m and 2.5-m telescope  observations,  Oleg Egorov for the help with preparing color illustrations, Aleksandrina Smirnova for the text improving and an anonymous reviewer for the comments and suggestions.    
\end{acknowledgements}


\begin{thebibliography}{10}
	\providecommand{\url}[1]{{#1}}
	\providecommand{\urlprefix}{URL }
	\expandafter\ifx\csname urlstyle\endcsname\relax
	\providecommand{\doi}[1]{DOI \discretionary{}{}{}#1}\else
	\providecommand{\doi}{DOI \discretionary{}{}{}\begingroup
		\urlstyle{rm}\Url}\fi
	
	\bibitem{Jockers1992}
	K.~{Jockers}, N.~{Thomas}, T.~{Bonev}, V.~{Ivanova}, V.~{Shkodrov}, Advances in
	Space Research \textbf{12}(8), 347 (1992).
	\newblock \doi{10.1016/0273-1177(92)90409-Q}
	
	\bibitem{Bland1998}
	J.~{Bland-Hawthorn}, D.H. {Jones}, \pasa \textbf{15}, 44 (1998).
	\newblock \doi{10.1071/AS98044}
	
	\bibitem{Jones2002}
	D.H. {Jones}, P.L. {Shopbell}, J.~{Bland-Hawthorn}, \mnras \textbf{329}, 759
	(2002).
	\newblock \doi{10.1046/j.1365-8711.2002.05001.x}
	
	\bibitem{Moiseev2010}
	A.~{Moiseev}, I.~{Karachentsev}, S.~{Kaisin}, \mnras \textbf{403}(4), 1849
	(2010).
	\newblock \doi{10.1111/j.1365-2966.2010.16254.x}
	
	\bibitem{MUSE}
	R.~{Bacon}, J.~{Vernet}, E.~{Borisova}, N.~{Bouch{\'e}}, J.~{Brinchmann},
	M.~{Carollo}, D.~{Carton}, J.~{Caruana}, S.~{Cerda}, T.~{Contini},
	M.~{Franx}, M.~{Girard}, A.~{Guerou}, N.~{Haddad}, G.~{Hau}, C.~{Herenz},
	J.C. {Herrera}, B.~{Husemann}, T.O. {Husser}, A.~{Jarno}, S.~{Kamann},
	D.~{Krajnovic}, S.~{Lilly}, V.~{Mainieri}, T.~{Martinsson}, R.~{Palsa},
	V.~{Patricio}, A.~{P{\'e}contal}, R.~{Pello}, L.~{Piqueras}, J.~{Richard},
	C.~{Sandin}, I.~{Schroetter}, F.~{Selman}, M.~{Shirazi}, A.~{Smette},
	K.~{Soto}, O.~{Streicher}, T.~{Urrutia}, P.~{Weilbacher}, L.~{Wisotzki},
	G.~{Zins}, The Messenger \textbf{157}, 13 (2014)
	
	\bibitem{MMTF}
	S.~{Veilleux}, B.J. {Weiner}, D.S.N. {Rupke}, M.~{McDonald}, C.~{Birk},
	J.~{Bland-Hawthorn}, A.~{Dressler}, T.~{Hare}, D.~{Osip}, C.~{Pietraszewski},
	S.N. {Vogel}, \aj \textbf{139}, 145 (2010).
	\newblock \doi{10.1088/0004-6256/139/1/145}
	
	\bibitem{Sugai2010}
	H.~{Sugai}, T.~{Hattori}, A.~{Kawai}, S.~{Ozaki}, T.~{Hayashi}, T.~{Ishigaki},
	M.~{Ishii}, H.~{Ohtani}, A.~{Shimono}, Y.~{Okita}, K.~{Matsubayashi},
	G.~{Kosugi}, M.~{Sasaki}, N.~{Takeyama}, \pasp \textbf{122}(887), 103 (2010).
	\newblock \doi{10.1086/650397}
	
	\bibitem{OSIRIS}
	J.J. {Gonz{\'a}lez}, J.~{Cepa}, J.I. {Gonz{\'a}lez-Serrano},
	M.~{S{\'a}nchez-Portal}, \mnras \textbf{443}, 3289 (2014).
	\newblock \doi{10.1093/mnras/stu1310}
	
	\bibitem{Jones2001}
	D.H. {Jones}, J.~{Bland-Hawthorn}, \apj \textbf{550}(2), 593 (2001).
	\newblock \doi{10.1086/319793}
	
	\bibitem{Shopbell1999}
	P.L. {Shopbell}, S.~{Veilleux}, J.~{Bland-Hawthorn}, \apjl \textbf{524}(2), L83
	(1999).
	\newblock \doi{10.1086/312311}
	
	\bibitem{Tinney1999}
	C.G. {Tinney}, A.J. {Tolley}, \mnras \textbf{304}(1), 119 (1999).
	\newblock \doi{10.1046/j.1365-8711.1999.02297.x}
	
	\bibitem{Courtes1960}
	G.~{Court{\`e}s}, Annales d'Astrophysique \textbf{23}, 115 (1960)
	
	\bibitem{Moiseev2002ifp}
	A.V. {Moiseev}, Bulletin of the Special Astrophysics Observatory \textbf{54},
	74 (2002)
	
	\bibitem{Jacquinot1954}
	P.~{Jacquinot}, Journal of the Optical Society of America (1917-1983)
	\textbf{44}(10), 761 (1954)
	
	\bibitem{SCORPIO2}
	V.L. {Afanasiev}, A.V. {Moiseev}, Baltic Astronomy \textbf{20}, 363 (2011)
	
	\bibitem{Dodonov2017}
	S.N. {Dodonov}, S.S. {Kotov}, T.A. {Movsesyan}, M.~{Gevorkyan}, Astrophysical
	Bulletin \textbf{72}(4), 473 (2017).
	\newblock \doi{10.1134/S1990341317040113}
	
	\bibitem{Oke1990}
	J.B. {Oke}, \aj \textbf{99}, 1621 (1990).
	\newblock \doi{10.1086/115444}
	
	\bibitem{Kartasheva1978}
	T.A. {Kartasheva}, N.M. {Chunakova}, Astrofizicheskie Issledovaniia Izvestiya
	Spetsial'noj Astrofizicheskoj Observatorii \textbf{10}, 44 (1978)
	
	\bibitem{Karachentsev2007}
	I.D. {Karachentsev}, S.S. {Kaisin}, \aj \textbf{133}, 1883 (2007).
	\newblock \doi{10.1086/512127}
	
	\bibitem{Kirsanova2019}
	M.S. {Kirsanova}, P.A. {Boley}, A.V. {Moiseev}, D.S. {Wiebe}, R.I. {Uklein},
	\mnras \textbf{497}(1), 1050 (2020).
	\newblock \doi{10.1093/mnras/staa2004}
	
	\bibitem{Kornilov2014}
	V.~{Kornilov}, B.~{Safonov}, M.~{Kornilov}, N.~{Shatsky}, O.~{Voziakova},
	S.~{Potanin}, I.~{Gorbunov}, V.~{Senik}, D.~{Cheryasov}, \pasp
	\textbf{126}(939), 482 (2014).
	\newblock \doi{10.1086/676648}
	
	\bibitem{Heckman1995}
	T.M. {Heckman}, M.~{Dahlem}, M.D. {Lehnert}, G.~{Fabbiano}, D.~{Gilmore}, W.H.
	{Waller}, \apj \textbf{448}, 98 (1995).
	\newblock \doi{10.1086/175944}
	
	\bibitem{Meaburn1989}
	J.~{Meaburn}, P.M. {Allan}, C.A. {Clayton}, A.P. {Marston}, M.J. {Whitehead},
	A.~{Pedlar}, \aap \textbf{208}, 17 (1989)
	
	\bibitem{Boroson1990}
	T.A. {Boroson}, \apj \textbf{360}, 465 (1990).
	\newblock \doi{10.1086/169136}
	
	\bibitem{Knese2020}
	E.D. {Knese}, W.C. {Keel}, G.~{Knese}, V.N. {Bennert}, A.~{Moiseev},
	A.~{Grokhovskaya}, S.N. {Dodonov}, \mnras \textbf{496}(2), 1035 (2020).
	\newblock \doi{10.1093/mnras/staa1510}
	
	\bibitem{Walter2002}
	F.~{Walter}, A.~{Weiss}, C.~{Martin}, N.~{Scoville}, \aj \textbf{123}, 225
	(2002).
	\newblock \doi{10.1086/324633}
	
	\bibitem{Keel2019}
	W.C. {Keel}, V.N. {Bennert}, A.~{Pancoast}, C.E. {Harris}, A.~{Nierenberg},
	S.D. {Chojnowski}, A.V. {Moiseev}, D.V. {Oparin}, C.J. {Lintott},
	K.~{Schawinski}, G.~{Mitchell}, C.~{Cornen}, \mnras \textbf{483}(4), 4847
	(2019).
	\newblock \doi{10.1093/mnras/sty3332}
	
	\bibitem{Silchenko2020}
	O.K. {Sil'chenko}, A.V. {Moiseev}, D.~{Oparin}, Astronomy Letters
	\textbf{accepted} (2020)
	
	\bibitem{Oparin2020}
	D.V. {Oparin}, O.V. {Egorov}, A.V. {Moiseev}, Astrophysical Bulletin
	\textbf{accepted} (2020)
	
\end{thebibliography}
\end{document}